\documentclass[reprint,eqsecnum,floats,aps,amsmath,amssymb,nofootinbib,prd,onecolumn, showpacs]{revtex4-2}

\usepackage{bm}
\usepackage{graphicx,physics}
\usepackage{graphicx}
\usepackage{amsmath,amssymb,mathtools,mathrsfs}
\usepackage{amsmath,amssymb,mathrsfs}
\usepackage{hyperref}
\usepackage{graphicx}
\usepackage{arydshln}
\usepackage{xcolor}
\usepackage{braket}
\usepackage{tensor}
\usepackage{enumitem,array,textcomp}
\usepackage[scr=boondoxo]{mathalpha}

\begin{document}

\title{Time-dependent scalings and Fock quantization of a massless scalar field in Kantowski-Sachs}
\author{Jer\'onimo Cortez}
\email{Electronic address: jacq@ciencias.unam.mx}
\affiliation{Departamento de F\'{\i}sica, Facultad de Ciencias, Universidad Nacional Aut\'onoma de M\'exico, Ciudad de M\'exico 04510, Mexico}

\author{Guillermo  A. Mena Marug\'an}
\email{Electronic address: mena@iem.cfmac.csic.es}
\affiliation{Instituto de Estructura de la Materia, IEM-CSIC, C/ Serrano 121, 28006 Madrid, Spain}

\author{Alvaro Torres-Caballeros}
\email{Electronic address: alvaro.torres@iem.cfmac.csic.es}
\affiliation{Instituto de Estructura de la Materia, IEM-CSIC, C/ Serrano 121, 28006 Madrid, Spain}

\author{Jos\'e Velhinho}
\email{Electronic address: jvelhi@ubi.pt}
\affiliation{Faculdade de Ci\^encias and FibEnTech-UBI, Universidade da Beira Interior, R. Marqu\^es D'\'Avila e Bolama, 6201-001 Covilh\~a, Portugal}

\def\reps{representations\ }
\def\l{\hat{l}}
\def\bl{\Tilde{l}}
\def\bll{\hat{\Tilde{l}}}
\def\bk{\Tilde{k}}
\def\z{z_{\bk\bll}}
\def\fhi{\mathring{\phi}_{n l m}}
\def\p{\mathring{\Pi}_{n l m}}
\def\a{\mathring{a}_{n l m}}
\def\aa{\mathring{a}_{n l m}^{*}}
\def\S{\mathscr{S}_{\l}}
\def\SS{\tilde{\mathscr{S}}_{\l}}
\def\SSS{\tilde{\mathfrak{s}}_{\bll}}
\def\n{\hat{n}}
\newcommand{\dtilde}[1]{\tilde{\raisebox{0pt}[0.85\height]{$\tilde{#1}$}}}
\def\DS{\mathscr{\Tilde{S}}_{\frac{n}{l}}}
\def\b{b_{\l}}
\def\x{\xi_{nl}}
\def\f{f_{nl}}
\def\g{g_{nl}}
\def\vaf{\vartheta_{nl}^{(f)}}
\def\vag{\vartheta_{nl}^{(g)}}
\def\s{s_{\l}}
\def\h{h_{nl}}
\def\bt{\Tilde{b}_{\l}}
\def\Hb{\mathscr{H}}

\begin{abstract}
We address the issue of inequivalent Fock representations in Quantum Field Theory in a curved homogenous and anisotropic background, namely Kantowski-Sachs spacetime. A family of unitarily equivalent Fock representations that are invariant under the spatial isometries and implement a unitary dynamics can be achieved by means of a field redefinition that consists of a specific anisotropic scaling of the field configuration and a linear transformation of its momentum. Remarkably, we show that this kind of field redefinition is in fact unique under our symmetry and unitary requirements. However, the physical properties of the Hamiltonian dynamics that one obtains in this way are not satisfactory, inasmuch as the action of the Hamiltonian on the corresponding particle states is ill defined. To construct a quantum theory without this problem, we need a further canonical transformation that is time- and mode-dependent and is not interpretable as an anisotropic scaling. The old and new Fock representations, nevertheless, are unitarily equivalent. The freedom that is introduced when allowing for this further canonical transformation can be fixed by demanding an asymptotic diagonalization of the Hamiltonian and a minimal absorption of dynamical phases. In this way, the choice of vacuum and the associated Fock representation are asymptotically determined.
\end{abstract}

\maketitle

\section{Introduction}\label{Intro}

The celebrated uniqueness of standard Quantum Mechanics is fundamentally rooted in the finite dimensionality of the phase space.  Indeed, for linear systems with finitely many degrees of freedom, the existence of different quantum representations is not a problem, because they are all unitarily equivalent to each other -- under appropriate requirements of continuity. This equivalence, stated in the well-known Stone-von Neumann theorem, is no longer valid when one moves to infinitely many degrees of freedom \cite{Wald, Simon}. The correspondence between the classical and quantum descriptions of a system is therefore not univocal and ambiguities aggravate the construction of the latter. Quantum field theories in flat spacetime deal with this kind of ambiguities in their Fock quantization by taking advantage of the background symmetries. Poincaré invariance plays here a key role, since it singles out a unique Fock representation, characterized precisely by the invariance of the vacuum state \cite{Wightman, Wald2}. The same idea can be extended to conformally flat backgrounds, such as de Sitter spacetimes in which the Bunch-Davies vacuum is defined e.g. by the identification with the Minkowski vacuum in an early-conformal-time limit \cite{Winitzky}. However, in more general cases it is not clear how to deal with the ambiguities in the quantization process. 

For a real scalar field, the freedom to select a suitable 1-particle Hilbert space from which the symmetric Fock space is to be constructed can be translated to the freedom of specifying an inner product on the complexification of the vector space of real, smooth solutions to the equations of motion. This, in turn, can be characterized by the so-called complex structures, namely a class of anti-selfadjoint operators $J$ with square equal to minus the identity and which endow the real vector space of solutions with a structure of complex vector space \cite{Wald, ReedSimon}. Hence, the choice of complex structure results in different and in (possible) inequivalent versions of the quantum theory \cite{AshtekarMagnon, Wald}. When the symmetries of the background are not enough to fix a single preferred complex structure, one is compelled to look for (physically plausible) extra criteria in order to eliminate the residual ambiguities. For instance, some proposals have been put forward over the years for the Fock quantization of a scalar field with an explicit time-dependent Hamiltonian --focusing primarily on isotropic cosmological backgrounds--, such as the algebraic approach \cite{Wald}, the classical-quantum energy requirement \cite{AshtekarMagnon}, the instantaneous lowest-energy state \cite{Winitzky}, the adiabatic vacuum \cite{Winitzky, Birrell}, the unitary quantum evolution \cite{CMV, CMOV}, etc. 

This last criterion, on which we concentrate our discussion, is based on preserving spatial background symmetries and requiring unitary quantum (Heisenberg) dynamics. For the nonstationary spacetimes that have been addressed, the program starts by transforming the field equations of the scalar field into equations of motion with a time-varying mass term, though in an effective static background. This is achieved by performing a time-depending canonical transformation, leading to field equations of the form:
\begin{equation} \label{mass}
\Ddot{\phi} -\Delta \phi +s(t)\phi = 0,
\end{equation}
where $\Delta$ is the Lapace-Beltrami operator of a static spatial metric, and $s(t)$ is  the effective mass function \cite{Seq}. In order to preserve the spatial background isometries, we can consider the class of invariant complex structures under the group of such spatial symmetries, and then study the dynamics of the linear combinations of the field and its conjugate momentum that are to be promoted to creation and annihilation variables \cite{Seq, PRD83, Laura1, Criteria, DreamTeam}. Indeed, it is possible to define a representation matrix of functions say $f$ and $g$, which relates the Laplace-Beltrami (LB) modes of the canonical fields with the creation and annihilation variables. These functions can depend on the LB eigenvalues (but do not mix them) and on time. Different pairs of functions $f$ and $g$ represent different complex structures and thus distinct representations \cite{DreamTeam}. On the other hand, as a way to impose a quantum unitary implementation of the dynamics, one is led to consider the evolution of the system as a Bogoliubov transformation with antidiagonal elements which must be square summable. As previously mentioned, this criterion successfully suppresses all unwanted ambiguities. In fact, it is possible to prove that even the first canonical transformation leading to \eqref{mass} is unique, in the sense that it is not possible to achieve a unitary implementation of the dynamics through any other transformation of the same type \cite{PRD83, Criteria}.

Given the rising interest in anisotropic cosmologies, the unitary dynamics criterion has also been tested in different scenarios of this kind, namely Bianchi I \cite{DreamTeam} and Kantowski-Sachs \cite{Prev} cosmologies. The interest in these anisotropic backgrounds goes in fact beyond cosmology. For instance, it has been shown that the interior geometry of nonrotating black holes can be foliated by a homogeneous but anisotropic class of space-like 3-manifolds and that this region is isometric to the Kantowski-Sachs cosmology \cite{AOS, AOSLet, AB}. Therefore, one can, for example, explore their quantum behavior \cite{Andres} by applying Loop Quantum Cosmology (LQC) techniques \cite{AP, EM}, a discipline inspired by Loop Quantum Gravity (LQG). LQG is a background-independent and nonperturbative approach to the quantization of General Relativity, and it is considered as one of the most solid candidates for a quantum theory of gravity \cite{AL, TH, RoVi}. 

With the aim at extending the above criterion to anisotropic cases, one is forced to consider a more general class of canonical transformations such that they can directly depend on the considered mode of the spatial LB operator, which makes the transformation nonlocal. At the end of the day, this is a common practice in the context of cosmological perturbations (see e.g. Ref. \cite{EM}). These nonlocal transformations nevertheless respect the existing spatial symmetries. Furthermore, as we will show in the next section, for the cases of Bianchi I  and Kantowski-Sachs spacetimes, the required mode dependence and therefore the involved locality breaking is marginal. Hence, the criterion still applies to these anisotropic cases. By means of particular transformations put forward in Refs. \cite{DreamTeam} and  \cite{Prev}, respectively for the Bianchi I  and Kantowski-Sachs cases, one can then obtain a unique family of unitarily equivalent Fock representations, a result that can be viewed as a natural generalization of the uniqueness of the quantum representation obtained in Refs.
 \cite{Seq, PRD83, Criteria} (for definiteness, see Sec. II of Ref. \cite{PRD83}), concerning in particular scalar fields in nonstationary isotropic spacetimes.

Nonetheless, it is still to be determined whether the time-dependent scalings involved in the aforementioned canonical transformation are unique or there exists any other possible transformation of this kind to obtain unitary quantum dynamics, thus introducing more inherent ambiguities. A result of uniqueness similar to the isotropic  case (generalizing in particular the results of Secs. III and IV of Ref. \cite{PRD83}) would be advantageous and beneficial. For example, if the Ashtekar-Olmedo-Singh blackhole model \cite{AOS} from LQC is likely to be considered with matter content, the determination of a unique family of Fock representations, and therefore of quantum operators, would place us in a privileged position for the quantization of the combined system. In the first part of the current article, we will show that the answer to this uniqueness issue is in the positive, just like in the isotropic case. More precisely, we will see that within the class of transformations of minimal nonlocality, i.e of the same type of those considered in Refs. \cite{DreamTeam} and  \cite{Prev}, the transformation is unique, and no further ambiguity is involved.

However, there is a drawback in adopting this minimal type of transformations, concerning in particular the quantum Hamiltonian. In fact, the straightforward quantization of the Hamiltonian does not possess some of the nice properties usually encountered in quantum field theory in flat spacetime. This issue is the subject of the second part of our work. We study the necessary conditions to obtain a well-defined action of the Hamiltonian on the vacuum, concluding  that  more general transformations are indeed required  for this purpose. There is thus a trade-off here: one can remain with minimal nonlocality, which gives a unique formulation leading to unitary dynamics, or live with slightly more general transformations, opening up the possibility for a standard and well understood quantization of the Hamiltonian, still in the context of unitary dynamics. The extra freedom introduced by this approach can, on the other hand, be addressed by means of a stronger requirement, already tested in simpler scenarios   \cite{BGT}, with the potential to remove all existing freedom and to select a preferred vacuum state.

The present work is organized as follows. In Sec. \ref{Preli} we present a summary of the unitary quantum evolution program applied to the Kantowski-Sachs background. In Sec. \ref{Det} other canonical transformations of the same type depending only on an anisotropic label and the background are considered for this anisotropic scenario. A detailed analysis of these transformations will show that the possible rescalings of the field are indeed unique, within the class of transformations considered in Ref. \cite{Prev}. In the first part of Sec. \ref{Prop} we study the action of the standard Hamiltonian onto the family of selected vacua. In this respect, transformations other than those considered in Sec. \ref{Det} are favored. Finally, in the last subsection, we introduce an ultraviolet Hamiltonian diagonalization which enables us to fix all the remaining freedom in the choice of representation. Moreover, this procedure determines a concrete splitting of the time evolution of the field between the background and a Heisenberg dynamics, along the lines of previous works \cite{DreamTeam, Prev}. This fact allows us to write the massless scalar field in terms of the creation and annihilation variables and the background. In addition, conclusions are presented in Sec. \ref{Co}. We adopt units such that the speed of light and Newton constant are equal to one, $c=G=1$.


\section{Preliminaries} \label{Preli}

We start by considering a massless real scalar field minimally coupled to a Kantowski-Sachs background, defined by the line element

\begin{equation}\label{2.1}
ds^2= -A^2(t) dt^2 + P^2(t)dr^2 + Q^2(t)(d\theta^2 + \sin^2\theta d\varphi^2 ).
\end{equation}
According to Ref. \cite{Prev}, the radial component $r$ has been compactified in a circle of period $2\pi$ to avoid infrared complications. Hence, the topology of the Cauchy surfaces is $S^1 \times \mathbb{S}^2$ (a three-handle), where $\theta$ and $\varphi$ are the usual coordinates of the sphere $\mathbb{S}^2$. One can now expand the real Klein-Gordon field $\Phi$ using a set $\{s_{nlm}\}$ of (normalized) real eigenfunctions of the LB operator associated with the  metric in the spatial manifold. This results in a description of the field $\Phi$ in terms of a discrete set of dynamical modes $\{\phi_{nlm}(t)\}$ which carry the time dependence of the field, namely
\begin{equation}\label{2.2}
\Phi (t,r,\theta,\varphi) = \sum_{nlm}\phi_{nlm}(t) s_{nlm}(r,\theta,\varphi), 
\end{equation}
where $n$ is summed over all integers, $l$ is summed over all positive integers and $m$ is summed over the integers in the interval $[-l,l]$. Similarly, the canonical conjugate momentum $\Pi$ can be expressed in the same manner, such that the nonvanishing Poisson brackets between phase space variables are those between the dynamical modes $\phi_{nlm}$ and the canonical conjugate modes of the momentum, $\Pi_{nlm}$. For further details, we refer the reader to Ref. \cite{Prev}.

Let us introduce a \textit{wavenumber} label $k\geq 0$ as 
\begin{equation}\label{2.3}
k= \sqrt{n^2+l(l+1)}. 
\end{equation}
Following Refs. \cite{Prev, DreamTeam} the zero mode, i.e. $k=0$, is left aside. It can always be treated by other means and does not affect the result of uniqueness (concerning an infinite number of modes). We also define a \textit{unit vector component} label as 
\begin{equation}\label{2.4}
\hat{l} = \frac{\sqrt{l(l+1)}}{k}.
\end{equation}
Then, the Hamiltonian of the scalar field is found to be 
\begin{equation}\label{2.5}
\mathbf{H}= \frac{1}{2} \sum_{nlm} (\Pi^2_{nlm} + P^2Q^4 W_{nl} \phi_{nlm}^2),
\end{equation}
where $W_{nl}$ is the eigenvalue of the LB operator:
\begin{equation}\label{2.6}
W_{nl} = \frac{k^2 \hspace{0.05cm} b_{\hat{l}}^2}{P^2Q^4}, \hspace{0.5cm} \text{with} \hspace{0.5cm} b_{\hat{l}}= Q^2 \sqrt{1+\hat{l}^2 \left( \frac{P^2}{Q^2}-1\right)},
\end{equation}
and we have chosen to describe the dynamics in harmonic time $\tau$, i.e. with the choice of lapse defined by the condition $A(t)dt = P(\tau)Q^2(\tau)d\tau$.

In order to apply the criteria commented in the Introduction and select a preferred family of unitary equivalent representations, we need to explore the evolution in the ultraviolet limit. For this, it is most convenient to seek a decoupling of the metric background functions $P(t)$ and $Q(t)$ from the labels $(n,l)$ in the eigenvalues $W_{nl}(\tau)$.

\subsection{Canonical transformation}\label{CanonicalTransformation}

We now perform the following canonical transformation:
\begin{equation}\label{2.7}
\Tilde{\phi}_{nlm} = \sqrt{b_{\l}} \ \phi_{nlm}, \hspace{1cm} \Tilde{\Pi}_{nlm} = \frac{1}{2} \frac{b_{\l}'}{b_{\l}^{3/2}} \ \phi_{nlm} + \frac{1}{\sqrt{b_{\l}}}\ \Pi_{nlm}.
\end{equation}
After this transformation, the Hamiltonian becomes
\begin{equation}\label{2.8}
\Tilde{\mathbf{H}}  = \Tilde{\sum_{nlm}} \frac{b_{\hat{l}}}{2} \left[ \tilde{\Pi}^2_{nlm} + \left( k^2 + s_{\l} \right) \tilde{\phi}_{nlm}^2 \right],
\end{equation}
where $s_{\l} = [3 (b_{\hat{l}}')^2 -2 b_{\hat{l}} b_{\hat{l}}'']/(4b_{\hat{l}}^4)$ is the mass function. 
In the above expressions, the prime denotes the derivative with respect to the harmonic time, and the tilde over the sum in Eq. \eqref{2.8}
indicates that the zero-mode, i.e $n=l=m=0$, has been left out. Note that the canonical transformation is time- and mode-dependent, a feature that is necessary for unitary quantum dynamics, as it is discussed in Ref. \cite{DreamTeam}. Nevertheless, the dependence on the mode, which makes the canonical transformation nonlocal, is limited, because it only comes from $\l$, which behaves as a unit vector in the space of labels $(n,l)$. Furthermore, the canonical transformation respects the spatial symmetries, since the transformation does not mix modes and it is independent of the label $m$. Finally, let us note that the resulting Hamiltonian corresponds to a scalar field in a static spacetime except for the global factor and the mass function which encode the anisotropic information of the system \cite{Laura1, Laura2, Criteria}.

It is possible to solve the dynamical equations in an asymptotic series (for large $k$) and then, by setting convenient initial conditions as discussed in Ref. \cite{DreamTeam}, the classical evolution from time $\tau_0$ to time $\tau$ can be expressed as the linear system
\begin{equation}\label{2.9}
	\begin{pmatrix}
            \tilde{\phi}_{nlm}\\
	    \tilde{\Pi}_{nlm}
	\end{pmatrix}_\tau = \mathcal{V}_{nl}(\tau,\tau_0) 
	\begin{pmatrix}
	\tilde{\phi}_{nlm}\\
	\tilde{\Pi}_{nlm}
	\end{pmatrix}_{\tau_0}, \qquad 
	\mathcal{V}_{nl}(\tau,\tau_0) = 
	\begin{pmatrix}
	\mathscr{R} \left\lbrace e^{i\Theta_{nl}} \right\rbrace & \frac{1}{k} \mathscr{I}\left\lbrace e^{i\Theta_{nl}}\right\rbrace \\
	- \frac{k^2+s_{\hat{l}}(\tau_0)}{k}\mathscr{I}\left\lbrace e^{i\Xi_{nl}}\right\rbrace & \mathscr{R} \left\lbrace e^{i\Xi_{nl}} \right\rbrace
	\end{pmatrix} ,
\end{equation}
where the symbols $\mathscr{R}$ and $\mathscr{I}$ denote the real and imaginary parts, respectively, and $\Theta_{nl}$ and $\Xi_{nl}$ are real functions of the asymptotic form $ k \eta_{\hat{l}}(\tau) + \mathcal{O}(k^{-1})$. Here, $\eta_{\l}$ is the conformal time defined as $\int_{\tau_0}^{\tau} d\bar{\tau} b_{\l}(\bar{\tau})$, and $\mathcal{O}(.)$ denotes the asymptotic order of its argument.

\subsection{Fock quantization with unitary dynamics}\label{FockQuant}

The next step is to choose a set of Fock representations of the canonical commutation relations for the scalar field (and its momentum) which respect the spatial symmetries of the Kantowski-Sachs spacetime and allow for a unitary implementation of the Heisenberg dynamics. The selection of this set of complex variables, which will be promoted to creation and annihilation operators in Fock space, can be attained by restricting ourselves to \textit{invariant} Fock representations, such that they do not mix modes and are independent of the label $m$. Thus, the matrix that relates the pair $(\tilde{\phi}_{nlm},\tilde{\Pi}_{nlm})$ with the complex variables $(a_{nlm},a^{*}_{nlm})$ of the representation is $m$-independent and block diagonal (we use the symbol $^{*}$ for complex conjugation). More explicitly,
\begin{equation}\label{2.10}
	\begin{pmatrix}
	a_{nlm} \\ a_{nlm}^{*}
	\end{pmatrix} = 
	\mathcal{F}_{nl}(\tau)
	\begin{pmatrix}
	\tilde{\phi}_{nlm} \\ \tilde{\Pi}_{nlm}
	\end{pmatrix}, \qquad \text{where} \quad \mathcal{F}_{nl}(\tau) = 
	\begin{pmatrix}
	f_{nl} (\tau)    & g_{nl}(\tau)  \\
	f_{nl}^{*} (\tau) & g_{nl}^{*}(\tau) 
	\end{pmatrix} .
\end{equation}
To guarantee that the new pair $(a_{nlm},a^{*}_{nlm})$ satisfy the standard Poisson algebra of creation and annihilation variables, it must further be required that
\begin{equation}\label{2.11}
f_{nl}(\tau) g^{*}_{nl}(\tau)  - g_{nl}(\tau) f^{*}_{nl}(\tau) =-i .
\end{equation}
Note that the functions $\f(\tau)$ and $\g(\tau)$ are allowed to be time-dependent. On top of the one coming from the canonical transformation \eqref{2.7}, this additional dependence introduces a splitting of the time dependence of the field and its momentum into a part that is assigned to the background and another part which is precisely the one that we want to be implemented as a unitary transformation. This splitting is in principle left unspecified and we only impose on it some natural requirements, which have been seen to be successful in other cases where it is convenient to separate the contribution from the time-varying background from the evolution of the system \cite{DreamTeam, Prev}. Note in particular that the mode dependence involved in the time-dependent transformations \eqref{2.10} is generally much stronger than the one coming from \eqref{2.7}, since the latter depends only on the combined parameter $\hat l$.

Dynamical evolution can be expressed in terms of our new variables as a set of Bogoliubov transformations of the form
\begin{equation} \label{2.12}
	\begin{aligned}
	\begin{pmatrix}
	a_{nlm} \\ a_{nlm}^{*}
	\end{pmatrix}_\tau = \mathcal{B}_{nl}(\tau,\tau_0) 
	\begin{pmatrix}
	a_{nlm} \\ a_{nlm}^{*}
	\end{pmatrix}_{\tau_0}, \hspace{0.6cm}
	\end{aligned}
 \text{where} \hspace{0.4cm} \begin{aligned}
	\mathcal{B}_{nl}(\tau,\tau_0) = 
	\begin{pmatrix}
	\alpha_{nl} (\tau,\tau_0)   & \beta_{nl}(\tau,\tau_0) \\
	\beta_{nl}^{*}(\tau,\tau_0) & \alpha_{nl}^{*} (\tau,\tau_0) 
	\end{pmatrix}
	= \mathcal{F}_{nl}(\tau)\mathcal{V}_{nl}(\tau,\tau_0)\mathcal{F}^{-1}_{nl}(\tau_0).
	\end{aligned}
\end{equation}
A general result \cite{Shale} shows that a necessary and sufficient condition for the transformation to be unitarily implementable is that it must be Hilbert-Schmidt, which in turn is equivalent to
\begin{equation}\label{2.13}
\tilde{\sum_{nlm}}  |\beta_{nl} (\tau,\tau_0)|^2 = \tilde{\sum_{nl}} (2l+1) |\beta_{nl} (\tau,\tau_0)|^2< \infty,
\end{equation}
for any time $\tau$. The factor $2l+1$ comes from the degeneracy in the label $m$. After a careful analysis, which can be found in Ref. \cite{Prev}, we conclude that the unitarity condition implies that 
\begin{equation} \label{2.14}
f_{nl}(\tau) = \sqrt{\frac{k}{2}} 
+ k \vartheta_{nl}^{(f)}(\tau), \qquad g_{nl}(\tau) = \frac{i}{\sqrt{2k}} 
+ \vartheta_{nl}^{(g)}(\tau),
\end{equation}
where the $\vartheta$-functions are subleading terms. Unitarity still constrains them, so that they must satisfy
\begin{equation} \label{2.15}
\tilde{\sum_{nlm}} k | \vartheta_{nl}^{(f)}(\tau) + i \vartheta_{nl}^{(g)}(\tau) |^2 = \tilde{\sum_{nl}} (2l+1)k | \vartheta_{nl}^{(f)}(\tau) + i \vartheta_{nl}^{(g)}(\tau) |^2< \infty, 
\end{equation}
for any time $\tau$. It was proven in Ref. \cite{Prev} that, by imposing this final condition and Eq. \eqref{2.11}, a set of transformations is selected such that its elements are connected among them by unitary relations. Therefore, the admissible Fock representations are all unitarily equivalent. In this sense, the quantization is unique. Among the representations in this family, there is one which will be helpful for the upcoming analysis. It is the \textit{massless representation} given by
\begin{equation} \label{2.16}
\breve{f}_{nl} = \sqrt{\frac{k}{2}}, \qquad \breve{g}_{nl} = \frac{i}{\sqrt{2k}}.
\end{equation}
It is a natural representation in the special case where the mass function $s_{\l}(\tau)$ vanishes for all $\tau$.

In fact, the emergence of the representation defined by relations \eqref{2.16} shows that no extra time dependence is required in Eq. \eqref{2.10} in order to attain unitary dynamics: the one arising from the canonical transformations \eqref{2.7} is already sufficient. On the other hand, conditions \eqref{2.14} and \eqref{2.15} imply that the requirement of unitary dynamics alone (together with the preservation of the spatial symmetries) is not sufficient to completely fix the aforementioned splitting of the field time dependence, at least if one allows for transformations as general as those in Eq. \eqref{2.10}.

In the next section, we will precisely show that this splitting is indeed fixed if, instead of the full generality of transformations \eqref{2.10}, we restrict our attention to transformations of the type \eqref{2.7}, i.e. with a time-dependent part which is only allowed to depend on $\hat l$. Nevertheless, we will return to general transformations \eqref{2.10} in Sec. \ref{Prop}, since they are seen to play an  important role in the construction of a well-defined quantum Hamiltonian.

\section{Determination of the scalings} \label{Det}

We have seen that the spatial isometries and a unitary Heisenberg evolution select a unique family of equivalent vacua for a massless scalar field in an anisotropic scenario, the Kantowski-Sachs background. Nevertheless, the question remains open whether the rescalings performed in Sec. \ref{CanonicalTransformation} are the only canonical ones that can lead to our results. To discuss this issue, in the following we will consider other linear canonical transformations of similar form, namely, transformations which change the field just by a scaling for each given $\l$ and add to the new momentum a linear contribution of the field:  
\begin{equation}\label{3.1}
    \begin{pmatrix}
        \fhi \\
        \p
    \end{pmatrix} = \mathcal{T}_{\l} 
    \begin{pmatrix}
        \tilde{\phi}_{nlm} \\
        \tilde{\Pi}_{nlm}
    \end{pmatrix}, \hspace{1cm} \mathcal{T}_{\l} = \begin{pmatrix}
        F_{\l} & 0 \\
        G_{\l} & \frac{1}{F_{\l}}
    \end{pmatrix},
\end{equation}
where $\mathcal{T}_{\l}(\tau)$ only depends on $\l$ and on the time $\tau$. Like in the isotropic case (see e.g. Secs. III and IV of Ref. \cite{PRD83}), the argument here is that, from the strict point of view of achieving unitary dynamics, there is no motivation for considering alternative transformations that are structurally different from \eqref{2.7}. One only needs to check if there are more transformations of the same type also leading to unitary dynamics. If that were the case, and if in particular different choices would lead to inequivalent quantum theories, then one would be introducing ambiguities in a process designed precisely to obviate them.  A careful analysis will show that in fact the only scaling of such a type which is compatible with an invariant Fock representation under the spatial symmetries and admits a unitary evolution is the one determined by Eqs. \eqref{2.7}. In order to prove this statement, we will show that our conditions restrict the transformation $\mathcal{T}_{\l}$ so that it can be considered trivial, i.e. $F_{\l} =1$ and $G_{\l}=0$, thus generalizing the results obtained in the isotropic context  \cite{PRD83}.

To start the analysis, let us first notice that the evolution of $\fhi$ and $\p$ is given by
\begin{equation} \label{3.2}
    \begin{aligned}
        \begin{pmatrix}
            \fhi \\
            \p
        \end{pmatrix}_\tau = \mathcal{T}_{\l} (\tau)\mathcal{V}_{nl}(\tau,\tau_0) \mathcal{T}^{-1}_{\l}(\tau_0) \begin{pmatrix}
            \fhi \\
            \p
        \end{pmatrix}_{\tau_0}.
    \end{aligned}
\end{equation}
We will use the simple massless representation as our reference representation,
\begin{equation} \label{3.3}
    \begin{pmatrix}
        \mathring{a}_{nlm} \\
        \mathring{a}_{nlm}^{*}
    \end{pmatrix} = \Breve{\mathcal{F}}_k \begin{pmatrix}
        \fhi \\ \p
    \end{pmatrix}, \hspace{1cm} \Breve{\mathcal{F}}_k = \frac{1}{\sqrt{2k}}\begin{pmatrix}
        k & i \\
        k & -i
    \end{pmatrix}.
\end{equation}
Combining Eqs. \eqref{3.2} and \eqref{3.3}, the corresponding creation and annihilation variables adopt the dynamics
\begin{equation} \label{3.4}
    \begin{pmatrix}
        \a \\
        \aa
    \end{pmatrix}_{\tau} = \mathring{U}(\tau,\tau_0) \begin{pmatrix}
        \a \\
        \aa
    \end{pmatrix}_{\tau_0}, \quad \text{where} \quad \mathring{U} = \begin{pmatrix}
        \mathring{\alpha}_{nl} & \mathring{\beta}_{nl}\\
        \mathring{\beta}_{nl} & \mathring{\alpha}_{nl}
    \end{pmatrix},
\end{equation}
and 
\begin{equation} \label{3.6}
    \begin{aligned}
        \mathring{U}{(\tau,\tau_0)} &= \breve{\mathcal{C}}(\tau)\  \breve{U}{(\tau,\tau_0)} \ \breve{\mathcal{C}}^{-1}(\tau_0), \\
        \Breve{\mathcal{C}}{(\tau)} &= \Breve{\mathcal{F}}_k \ \mathcal{T}_{\l}{(\tau)} \ \Breve{\mathcal{F}}_k^{-1},\\
        \breve{U} {(\tau,\tau_0)}&= \Breve{\mathcal{F}}_k\ \mathcal{V}_{nl}(\tau,\tau_0) \ \Breve{\mathcal{F}}_k^{-1} . 
    \end{aligned}
\end{equation}
In this equation, $ \Breve{\mathcal{C}}$ has the form 
\begin{equation}\label{3.7}
     \Breve{\mathcal{C}} = \begin{pmatrix}
         F_{\l}^{+}+ \frac{iG_{\l}}{2k} & F_{\l}^{-} + \frac{iG_{\l}}{2k} \\
         F_{\l}^{-}- i \frac{iG_{\l}}{2k} & F_{\l}^{+}- \frac{iG_{\l}}{2k}
     \end{pmatrix}, \text{ where} \hspace{0.4cm} 2F_{\l}^{\pm} = F_{\l} \pm \frac{1}{F_{\l}}.
\end{equation}
To allow for other representations in the new dynamics, let us now consider an invariant, but otherwise generic complex structure, and recall that different complex structures characterize different Fock representations. They are defined by time-independent matrices of the form
\begin{equation}\label{3.8}
    \mathring{J}_{nl} = \begin{pmatrix}
        \delta_{nl} & \lambda_{nl} \\
        \lambda^{*}_{nl}& \delta^{*}_{nl}
    \end{pmatrix}, \text{ with} \hspace{0.4cm} |\delta_{nl}|^2-|\lambda_{nl}|^2=1.
\end{equation}
In the chosen representation, we want the dynamics to be unitary. The dynamical evolution of the creation and annihilation variables amounts to a Bogoliubov transformation,  $ \mathring{U}^{J}$, which is just 
\begin{equation} \label{3.9}
    \mathring{U}^{J} =  \mathring{J}_{_{nl}}^{-1}  \mathring{U} \ \mathring{J}_{_{nl}}  = \begin{pmatrix}
        \mathring{\alpha}^{J}_{nl} & \mathring{\beta}^{J}_{nl}\\
        \mathring{\beta}^{J*}_{nl} & \mathring{\alpha}^{J*}_{nl}
    \end{pmatrix},
\end{equation}
where $\mathring{J}_{nl}$ is the matrix of the complex structure introduced above. 

Note that the matrices $\mathring{U}^{J}$ are just special forms of the matrices $\mathcal{B}_{nl}$ appearing in Eq. \eqref{2.12}, coming in particular from matrices $\mathcal{F}_{nl}$ of the form 
\begin{equation} \label{3.z}
    \mathcal{F}_{nl} (\tau)=  \mathring{J}_{_{nl}}^{-1}  \Breve{\mathcal{F}}_k \mathcal{T}_{\l} (\tau).
		 \end{equation}
We are thus dealing with a particular type of transformations \eqref{2.10} where the mode dependence of the time-varying part is restricted to be of the type \eqref{3.1}, in order to preserve the structure of the transformations \eqref{2.7}.

Unitarity amounts then to the requirement
\begin{equation} \label{3.10}
    \tilde{\sum_{n l m}} |\mathring{\beta}^{J}_{nl}|^2  < \infty ,
\end{equation}
where, according to Eq. \eqref{3.z}, the entries $f_{nl}$ and $g_{nl}$ of the matrices $\mathcal{F}_{nl}$ now take the form
\begin{equation}\label{3.16}
    \begin{aligned}
        f_{nl} &= \frac{1}{\sqrt{2k}} \left( k \hspace{0.1cm}  F_{\l} \hspace{0.1cm} [ \delta^{*}_{nl}-\lambda_{nl}] + i\hspace{0.1cm} G_{\l} \hspace{0.1cm}  [\delta^{*}_{nl}+\lambda_{nl}] \right) , \\
        g_{nl} &= \frac{i}{\sqrt{2k}} \frac{[\delta^{*}_{nl}+\lambda_{nl}]}{F_{\l}}.
    \end{aligned}
\end{equation}
The unitarity condition \eqref{2.15} then leads to the square summability of the following expression:
\begin{equation} \label{3.17}
   \sqrt{k}\left( \vartheta^{(f)}_{nl}(\tau)+i \vartheta^{(g)}_{nl}(\tau)\right)= \frac{1}{\sqrt{2}} \left(2 \delta^{*}_{nl} F_{\l}^{-} - 2\lambda_{nl} F_{\l}^{+} + \frac{i}{k} G_{\l} [\delta^{*}_{nl}+\lambda_{nl}]\right).
\end{equation}
To prepare it for our subsequent analysis, let us divide the above expression by $\sqrt{2}\delta^{*}_{nl}$. 
Recall that, for every $\delta_{nl}$ and $\lambda_{nl}$ $\in \mathbb{C}$, they satisfy 
$|\delta_{nl}|^2 = 1 + |\lambda_{nl}|^2\geq 1$, and therefore this division does not affect the summability. Thus, for a transformation of the type \eqref{3.1} to be compatible with unitary dynamics in a representation defined by an invariant complex structure such as the one given by Eqs.  \eqref{3.8}, it is necessary that
\begin{equation}\label{3.18}
   \tilde{\sum_{nlm}} \Bigg|  F_{\l}^{-} - \frac{\lambda_{nl}}{\delta^{*}_{nl}} F_{\l}^{+} + \frac{i}{2k} G_{\l} \left[1+\frac{\lambda_{nl}}{\delta^{*}_{nl}}\right] \Bigg|^2 < \infty.
\end{equation}
Clearly, square summability must be satisfied on any given subsequence. In the following, we will take advantage of two special types of sequences to obtain our result. First, we will show that $F_{\l}(\tau)$ is in fact time-independent for any given $\l$, and can therefore be set to $F_{\l}=1$, $\forall \l$, by means of a single time-independent canonical transformation.  Afterwards, a natural regularity condition will lead to  
$G_{\l}=0$.

\subsection{Function $F_{\l}$}\label{SIIIA}

Let us start by showing that, for any given $\l$, there exists an infinite sequence $\mathscr{S}_{\l}$  composed of triplets $\lbrace (n_i,l_i,m_i) \rbrace_i$, $i\in\mathbb{N}$, with $m_i= l_i$, such that the values of $\hat{l}$, and therefore of $F_{\l}$ and $G_{\l}$, are constant on $\S$, while $k$ grows unboundedly. 

For given $\l$, let $n_1$ and $l_1$ be such that 
$l_1(l_1+1) [n_1^2+l_1(l_1+1)]^{-1}=\l^{2}$, or equivalently  $n_1^2 [l_1(l_1+1)]^{-1}=\l^{-2}-1$. 
The sequence $\S$ is then generated by the following recurrence relation
\begin{equation}\label{3.13}
    \begin{aligned}
        n_{i+1} =2n_i(2l_i+1) \\
        l_{i+1}=4l_i(l_i+1),  
    \end{aligned}
\end{equation}
$i=1,2,\cdots$, together with the condition  $m_i= l_i, \forall i$.
Note that the sequence of values of $k$ is given by the recurrence relation $k_{i+1}^2=4(2l_i+1)^2 k_i^2$, and therefore the sequence $\lbrace k_i \rbrace_{i\in\mathbb{N}}$
is unbounded over  $\S$, for any set of initial values $(n_1,l_1)$ (except $n_1=l_1=0$).


We can now go back to condition \eqref{3.18}. It implies in particular that the expression of which the square norm is being summed must tend to zero in any infinite subsequence. Then, for any fixed $\l$, we can take the corresponding sequence $\S$, to conclude  that the resulting series must have a vanishing limit on $\S$, i.e.
\begin{equation}\label{3.z2}
     F_{\l}^{-} - \frac{\lambda_{nl}}{\delta^{*}_{nl}} F_{\l}^{+} + \frac{i}{2k} G_{\l} \left[1+\frac{\lambda_{nl}}{\delta^{*}_{nl}}\right] \overset{\S}{\longrightarrow} 0,
\end{equation}
where the arrow denotes the behaviour over the sequence $\S$.

However,  the term $\left[1+\frac{\lambda_{nl}}{\delta^{*}_{nl}}\right]$ is bounded (again because $|\delta_{nl}|^2 = 1 + |\lambda_{nl}|^2$), $G_{\l}$ is constant in $\S$ (though time-dependent) and $k$ grows unboundedly in $\S$. Thus,
\begin{equation}\label{3.19}
    \frac{i}{2k} G_{\l} \left[1+\frac{\lambda_{nl}}{\delta^{*}_{nl}}\right] \overset{\S}{\longrightarrow} 0.
\end{equation}
Moreover, $F_{\l}^{\pm}$ are constant in $\mathscr{S}_{\l}$. The only possibility left so that \eqref{3.z2} is satisfied is that the limit of $\lambda_{nl}/\delta_{nl}^{*}$ in $\mathscr{S}_{\l}$ exists and equals 
\begin{equation}\label{3.20}
    \lim_{\S} \frac{\lambda_{nl}}{\delta^{*}_{nl}} = \frac{F_{\l}^{-}}{F_{\l}^{+}}= \frac{F^2_{\l} -1}{F^2_{\l} +1}.
\end{equation}
Since neither $\lambda_{nl}$ nor $\delta_{nl}$ depend on time, the scaling $F_{\l}$ does not depend on $\tau$. Actually, this conclusion is valid $\forall \l$. The scalings $F_{\l}$ can then be removed using a time-independent canonical transformation and, once this is done, we can take $F_{\l}=1$ (see Refs. \cite{CMV,PRD83} for details). 

Let us note that condition \eqref{3.18} with $F_{\l}=1$ still applies.  In the next subsection, we will explore the properties of another type of special sequences, to show that the unitarity condition \eqref{3.18} requires $G_{\l}=0$ for all $\l$.

\subsection{Function $G_{\l}$}

Let us start by noting that fixing $\l$ is equivalent to fixing $\n=n [l(l+1)]^{-1/2}$, because 
\begin{equation}\label{3.22}
    \l = \left( \frac{l(l+1)}{n^2+l(l+1)} \right)^{1/2} =\left( \frac{1}{\n^2+1} \right)^{1/2}.
\end{equation}
Consider $n,l \in \mathbb{N}$ such that its $gcd(n,l)=1$ (with $gcd$ standing for greatest common divisor), i.e. they form an irreducible fraction. Then, take the sequence $\DS$ formed by the triplets $\{ n_j,l_j,m_r\}_{(j,r)}$ where $n_j=j n$, $l_j= j l$, and $m_r\in \{ - l_j, -l_j+1,..., l_j-1,l_j\}$, with $j\in \mathbb{N}$ and $r=1,..., 2l_j +1$. Note that $\n_{j}=n_j [l_j(l_j+1)]^{-1/2}$ is such that
\begin{equation}\label{3.23}
    \n_{j}= \frac{jn}{\sqrt{(jl)(jl+1)}},
\end{equation}
and therefore the sequence $\{\n_{j}\}$  clearly converges to $n/l$ when $j$ tends to infinity. Since $n$ and $l$ are arbitrary (provided $gcd(n,l)=1$), we can affirm that there are points in the sequences $\{\n_{j}\}$ arbitrarily close to any irreducible fraction. Since the set of irreducible fractions is basically the set of rational numbers $\mathbb{Q}$, which is dense, so is the set of all possible values of $\{\n_j\}$. In addition, note that given a triplet $(n,l,m)$, there exists a set $\DS$ which contains it, which is characterized by the corresponding irreducible fraction $n/l$. The only special cases would be $l=0$ or $n=0$ (not both simultaneously, since we do not include zero modes). Since they specify a finite number of modes, they are not relevant to our arguments, related to the field-like behavior.

Note also that the relation $\l=(\n^2+1)^{-1/2}$ can be extended to all real values of $\n$, in a way that is $C^{\infty}$. The image of $n/l$ is given by $\l=l/N$, where we have defined $N=(n^2+l^2)^{1/2}$. 

Strictly speaking, the set of irreducible fractions $\mathbb{Q}$ is not covered by the set of possible values of $\n_j$ (see Eq. \eqref{3.23} with $j=1$, for instance), and thus the corresponding values of $\l$ do not run over the whole of the interval $(0,1)$ (and henceforth over $[0,1]$). Nonetheless, by taking functions $G_{\l}$ that are smooth, we can extend their domain in $\l$ to include all the points in $[0,1]$. Hence, from now on, we assume that the function $G_{\l}$ is regular in $\l$, in the sense that it does not vary wildly between any $\l_{j}=(\n_j^2 +1)^{-1/2}$ and $\l_{j+1}=(\n_{j+1}^2 +1)^{-1/2}$. More concretely, here we take $G_{\l}(\tau)$ as the restriction to (the union of the sets) $\{ \l_j=(\n_{j}^2+1)^{-1/2} \}$ of a Lipschitz continuous function $G(\l,\tau): [0,1]\times \mathbb{I} \to \mathbb{R}$, where $\mathbb{I}$ is the (compact) time interval under consideration. Hence, there exists $M$ such that $\forall \l_j$, $\l_{j+1}$, we have $|G_{\l_j} (\tau)-G_{\l_{j+1}}(\tau)|\leq M |\l_j-\l_{j+1}|$ for all $\tau \in \mathbb{I}$. This ensures that the variation of $G_{\l}(\tau)$ for all times in $\mathbb{I}$ is of the same order as that of $\l$ in our sequence. For large $j$, it is not difficult to check that this last variation is of order $1/j$, which is of order $k^{-1}$, since $k_j=[n_j^2+l_j(l_j+1)]^{1/2}$. Subdominant corrections to this relation are of order $k^{-2}$.

A relevant property of the sequences $\DS$ is the following. Let $\mu(n,l,m)=\mu(n,l)$ be a function that does not depend on $m$. Then,
the sum over $\DS$ becomes
\begin{equation}\label{3.24}
    \sum_{\DS} \mu(n,l) = \sum_{j=1}^{\infty} (2 l_{j} +1) \ \mu(n_{j},l_{j}) \approx  \sum_{j=1}^{\infty} j\ \mu(n_{j},l_{j}),
\end{equation}
where the symbol $\approx$ denotes a similar asymptotic order for large values of $j$, disregarding (nonvanishing) global constant factors. 

Consider again condition \eqref{3.18}, now particularized to the case of a unit function $F_{\l}$. Summability must also hold for any infinite subsequence of triples $(n,l,m)$, and in particular for sequences of the type $\DS$. According to our previous discussion, unitarity then requires\footnote{At this point, it is worth commenting on some subtleties concerning our asymptotic analysis. Square summability of the sequence \eqref{3.17} as expressed in Eq. \eqref{2.15} was derived from Eqs. (5.6)-(5.8) in our previous work on Kantowski-Sachs \cite{Prev} by assuming that the time dependence of our change of variables cannot trivialize the dynamics of the system. These dynamics depend on the label $\l$ of the modes through two terms: the time-dependent mass $s_{\l}$, and the definition of the conformal time $\eta_{\l}$ where the function $b_{\l}$ appears. In our asymptotics, we used that $s_{\l}$ remains bounded over all modes in order to obtain the behavior (5.8) of Ref. \cite{Prev}. This shows a functional dependence on the imaginary exponentials of $k\eta_{\l}$. It is this functional dependence that we assumed cannot be compensated with the time dependence of our definition of basic variables. We can gain some insight into what this assumption means for the subsequence of modes that we are considering. For smooth metric functions, the function $b_{\l}$ has Lipschitz continuous derivatives, so that $b_{\l}$ at different values of $\l$ in $\DS$ can be expressed as the function at the limiting point $l/N$, plus the derivative of the function with respect to $\l$ evaluated at $l/N$ multiplied by the difference $(\l-l/N)$, plus something of order $(\l-l/N)^2$. On the other hand, we know that $(\l-l/N)$ is of order $1/j$ in our sequence, and hence of order $1/k$. In total, up to terms of order $k^{-2}$, we can write $\eta_{\l}=\eta_{l/N}+ \xi_{l/N}/k$, where $\xi_{l/N}$ is a function of time proportional to the integral over time of the derivative of $b_{\l}$ with respect to $\l$, at $\l=l/N$. Notice that $\xi_{l/N}$ does not depend on the mode in $\DS$, but only on the fixed value $l/N$. Therefore, up to irrelevant subdominant terms, the imaginary exponentials of $k\eta_{\l}$ become $e^{\pm i k \eta_{l/N}}$ multiplied by a time-dependent phase $e^{\pm i \xi_{l/N}(\tau)}$ that is mode-independent in our sequence. Then, the nontrivialization of the dynamics in our sequence amounts to the functional independence with respect to the conformal Fourier modes $e^{\pm i k \eta_{l/N}}$. The phase $e^{\pm i \xi_{l/N}(\tau)}$ does not affect the square summability of the factors multiplying these conformal Fourier modes.}
\begin{equation}\label{3.25}
\sum_{\DS} \Bigg| -\frac{\lambda_{nl}}{\delta^{*}_{nl}} + \frac{i}{2k} G_{\l} \left[ 1+ \frac{\lambda_{nl}}{\delta^{*}_{nl}}\right] \Bigg|^2 < \infty.
\end{equation}
Furthermore, summability in $\DS$ must also hold when we divide the summands by $k$ (since $k\geq 1$ in our sequence), leading to
\begin{equation}\label{3.26}
    \sum_{\DS} \Bigg| -\frac{1}{k}\frac{\lambda_{nl}}{\delta^{*}_{nl}} + \frac{i}{2k^2} G_{\l} \left[ 1+ \frac{\lambda_{nl}}{\delta^{*}_{nl}}\right] \Bigg|^2 < \infty.
\end{equation}
Taking into account that the values of $\l$ converge to $l/N$ over $\DS$, we conclude that $G_{\l}(\tau)$ converges to $G(l/N,\tau)$ (recall that the Lipschitz condition on $G(\l,\tau)$ ensures that the corrections to the limit are of order $k^{-1}$, which is the order of the corrections to $\l=l/N$ in our sequence). Given the boundness of the term $1+\lambda_{nl}/\delta^{*}_{nl}$, we then obtain the behavior
\begin{equation}\label{3.27}
    \sum_{\DS} \Bigg| \frac{1}{k^2} G_{\l} \left(1+\frac{\lambda_{nl}}{\delta^{*}_{nl}}\right) \Bigg|^2 
    \approx  \sum_{j=1}^{\infty}G^2(l/N) \frac{j}{k_j^4} 
    \approx G^2(l/N) \sum_{k\in\{k_j\}}  \frac{1}{k^3},
\end{equation}
which shows that the second term in the sum \eqref{3.26} is square summable. But this implies that the first term must be square summable as well. Using this result in condition \eqref{3.25}, together with the boundedness of $G(l/N,\tau)$ in $\mathbb{I}$ (and restoring the time dependence in our notation), it follows that we must have, $\forall \tau$,
\begin{equation}\label{3.28}
    \sum_{\DS} \Bigg| -\frac{\lambda_{nl}}{\delta^{*}_{nl}} + \frac{i}{2k} G(l/N,\tau) \Bigg|^2<\infty.
\end{equation}
However, unless $G(l/N,\tau)$ does not really depend on time (in which case we can remove it by means of a time-independent canonical transformation
\cite{CMV,PRD83}), it follows that both terms must be square summable, because their difference cannot cancel, since one term depends on time and the other does not. We are thus led to conclude that
\begin{equation}\label{3.29}
    \sum_{\DS} \Bigg| \frac{1}{k} G(l/N,\tau) \Bigg|^2 = |G(l/N,\tau)|^2  \sum_{\DS} \frac{1}{k^2} < \infty
\end{equation}
in $\mathbb{I}$. But, from Eq. \eqref{3.24}, this sum behaves as $\sum_{k\in\{k_j\}} (1/k)$, which diverges. So, the unitarity condition is impossible to fulfil unless $G(\l,\tau)=0$ for all times in our interval and $\l$ in the image of any $\n=n/l \in \mathbb{Q}$. Since $\mathbb{Q}$ is dense and $G$ is Lipschitz continuous, it follows that $G$ is identically zero.

\section{Properties of the Hamiltonian} \label{Prop}

As previously mentioned, there is great freedom in the transformation \eqref{2.10} performed in Sec. \ref{FockQuant} to pass from the canonical pair $(\tilde{\phi}_{nlm}, \tilde{\Pi}_{nlm})$ to annihilation and creation variables $(a_{nlm},a^{*}_{nlm})$. In general, the functions $f_{nl}$ and $g_{nl}$ in this transformation introduce a further splitting of the time evolution of the system, assigning part of it to the background. At this stage, it is unclear that one needs to resort to these kinds of generally time-dependent transformations beyond the anisotropic scalings that we have already discussed and proven to be unique. To clarify this issue, let us now consider exclusively the family of unitarily equivalent representations which are obtained from the field redefinition determined in the previous section by permitting only time-independent canonical transformations. We will study the physical properties of their Hamiltonian. 

\subsection{Time-independent functions $f_{nl}$ and $g_{nl}$ }

We now restrict the functions $f_{nl}$ and $g_{nl}$ in Eqs. \eqref{2.10} to be independent of time. As such, their only role is to parametrize different representations of the canonical commutation relations compatible with the spatial symmetries. Moreover, any such two representations are unitarily equivalent, if they allow a unitary implementation of the dynamics of the canonical pair $(\tilde{\phi}_{nlm}, \tilde{\Pi}_{nlm})$ \cite{Prev}. Under the canonical transformation that they determine, the Hamiltonian can be expressed as
\begin{equation}\label{4.1}
    \begin{aligned}
    H = \tilde{\sum_{nlm}} \left( - \frac{b_{\l}}{2} \right) &\Bigl\{ a^{*}_{nlm}a^{*}_{nlm} \left[f_{nl}^2 + g_{nl}^2 \left(k^2 + s_{\l} \right) \right] + a_{nlm}a_{nlm} \left[f_{nl}^{*2} + g_{nl}^{*2} \left(k^2 + s_{\l} \right) \right] \\ &-2a^{*}_{nlm}a_{nlm} \left[ |f_{nl}|^2+ |g_{nl}|^2 \left(k^2 + s_{\l} \right) \right]\Bigr\}.
\end{aligned} 
\end{equation}
We again use the tilde over the sum to denote that it extends over nonzero modes. 

Let us quantize this Hamiltonian in the standard way, i.e. promoting the variables $(a_{nlm},a^{*}_{nlm})$ to annihilation and creation operators, and adopting normal ordering. As usual, the zero particle state $\ket{0}$ of the associated Fock space is henceforth called the vacuum state. For this quantum Hamiltonian $\hat{H}$ to have a well-defined action on the vacuum, the image $\hat{H}\ket{0}$ should have a finite norm, namely
\begin{equation} \label{4.2}
\bra{0} \hat{H}^\dagger \hat{H}\ket{0} = \tilde{\sum_{nlm}} \frac{b^2_{\l}}{2} \ |f_{nl}^2 + g_{nl}^2(k^2+s_{\l})|^2 < \infty.
\end{equation}
On the other hand, from Eq. \eqref{2.11} we get that the imaginary part of $f^{*}_{nl}g_{nl}$ is 1/2. Hence,
\begin{equation} \label{4.3}
    |f_{nl}| \ |g_{nl}| \geq \frac{1}{2}.
\end{equation}
Using this result, if $f_{nl}$ and $g_{nl}$ are of the same asymptotic order when $k\rightarrow \infty$, then they both need to be of order $\mathcal{O}(k^0)$. However, this would imply that the summands in Eq. \eqref{4.2} do not tend to zero as $k\to \infty$, and consequently the partial sums for a given $\l$ (or $\n$) would diverge. Suppose that the asymptotic order of $f_{nl}$ is then smaller than that of $g_{nl}$. Taking into account Eq. $\eqref{4.3}$, the order of $g_{nl}$ needs to be larger than $\mathcal{O}(k^0)$. Nevertheless, that would imply again that Eq. \eqref{4.2} diverges. Thus, the only possibility is that the order of $f_{nl}$ be greater than the order of $g_{nl}$, and that both terms inside the norm cancel each other at dominant order. This means that  $k^2g_{nl}^2 = -f_{nl}^2$ up to subdominant terms, and therefore 
\begin{equation}\label{4.4}
     k g_{nl} = if_{nl} + \xi_{nl}.
\end{equation}
Here, the order of $f_{nl}$ is $\mathcal{O}\left(k^{1/2}\right)$ and the order of $g_{nl}$ is $\mathcal{O}\left(k^{-1/2}\right)$, owing to Eq. \eqref{4.4}, and $\xi_{nl} = o\left(k^{1/2} \right)$, where the symbol $o(.)$ denotes asymptotically negligible with respect to its argument. We have taken the dominant behavior $i f_{nl}$ rather than its negative, which would imply, together with Eq. \eqref{2.11}, that $2|f_{nl}|^2/k=-1+o(k^{0})$, something impossible. Moreover, from the same Eq. \eqref{2.11} and Eq. \eqref{4.4}, we find
\begin{equation}\label{4.5}
    |f_{nl}|^2 = \frac{k}{2} + o(k).
\end{equation}
Therefore, with a convenient choice of an irrelevant global phase, we arrive at functions $\f$ and $\g$ of our previous form, i.e.
\begin{equation} \label{4.6}
\begin{aligned}
    f_{nl} = \sqrt{\frac{k}{2}} + k \vartheta^{(f)}_{nl}  \\
    g_{nl} = \frac{i}{\sqrt{2k}} + \vartheta^{(g)}_{nl}, 
    \end{aligned}
\end{equation}
where the order of $\vartheta^{(f)}_{nl}$ and $\vartheta^{(g)}_{nl}$ must be $o(k^{-1/2})$. 
Hence, we can write $g_{nl}$ as
\begin{equation}\label{4.7}
    g_{nl} = \frac{i}{k}f_{nl} -i \left( \vartheta^{(f)}_{nl} +i \vartheta^{(g)}_{nl} \right).
\end{equation}
Substituting $g_{nl}$ into Eq. \eqref{4.2} and recalling that $b_{\hat{l}}$ is bounded in our set of modes, we obtain 
\begin{equation} \label{4.8}
   \tilde{\sum_{nlm}} \Bigg| \left[2f_{nl} k \left( \vartheta^{(f)}_{nl} +i \vartheta^{(g)}_{nl} \right) - k^2 \left( \vartheta^{(f)}_{nl} +i \vartheta^{(g)}_{nl} \right)^2 \right] \left(1 + \frac{s_{\l}}{k^2} \right)- \frac{s_{\l}}{k^2} f^2_{nl} \Bigg|^2 < \infty.
\end{equation}
The square summability must also hold over the subsequence of modes $\DS$. Substituting then $f_{nl}$ into our equation yields
\begin{equation}\label{4.9}
\begin{aligned}
   \sum_{\DS} \Bigg| \left( \vartheta^{(f)}_{nl} +i \vartheta^{(g)}_{nl} \right)          \left[\sqrt{2k^3} - k^2 \left( \vartheta^{(f)}_{nl} +i \vartheta^{(g)}_{nl} \right) + 2k^2  \vartheta^{(f)}_{nl}  \right] \left(1 + \frac{s_{\l}}{k^2} \right)  - \frac{s_{\l}}{k^2} \left( \sqrt{\frac{k}{2}} + k \vartheta^{(f)}_{nl} \right)^2  \Bigg|^2 < \infty.
   \end{aligned}
\end{equation}
Recall that, in our sequence, $s_{\l}$ tends to $s_{l/N}$ up to subdominant terms in $k$ (because the difference between $\l$ and $l/N$ is of this order and because $s_{\l}$ is Lipschitz continuous from its definition, for smooth metric functions). In addition, we notice that
$k^2 ( \vartheta^{(f)}_{nl} +i \vartheta^{(g)}_{nl} )$ and the term $2k^2  \vartheta^{(f)}_{nl}$ are $o(k^{3/2})$. Hence, at dominant order, we only have to keep the terms with  $\sqrt{2k^3}$ and $\sqrt{k/2}$ in the first square brackets and the last round brackets, respectively. We get
\begin{equation}\label{4.10}
    \sum_{\DS} \Bigg|\left( \vartheta^{(f)}_{nl} +i \vartheta^{(g)}_{nl} \right) \sqrt{2k^3} - \frac{s_{l/N}}{2k}\Bigg|^2 < \infty.
\end{equation}
The degeneracy in $m$ introduces a factor growing like $k$ when this part of the sum is made. Then, it is not difficult to see that the only way for the total sum to converge is that the contribution proportional to $s_{l/N}$ is cancelled by the other terms. However, this is impossible because $s_{l/N}(\tau)$ is a nonvanishing function of time, while the linear combination of the functions $\vartheta^{(f)}_{nl}$ and $\vartheta^{(g)}_{nl}$ that appears in the formula is time-independent. Thus, the action of the Hamiltonian is ill defined at all times on the vacuum, and therefore in all the associated particle states. 
Note that this is true for any possible representation (preserving the spatial symmetries) that allows a unitary implementation of the dynamics corresponding to the pair $(\tilde{\phi}_{nlm}, \tilde{\Pi}_{nlm})$ if the functions $f_{nl}$ and $g_{nl}$ are constant in time.

A priori, the issue of an ill-defined Hamiltonian action might be addressed by different approaches, e.g. renormalization. However, here we will consider a totally different route that will benefit from the presence of a dynamic background. As we mentioned in the Introduction, one of the aims of the present work is to prepare the Hamiltonian description for the case of a hybrid quantization in Loop Quantum Cosmology where, in addition to the Fock quantization of the field, the background itself is quantized using Loop techniques. In this context, renormalization would not seem natural, since we could not attribute the presence of divergences to the consideration of the background geometry as a classical entity. Instead, we will allow for time-dependent representation functions via their possible dependence on the background. We will then study the necessary conditions to obtain a well-defined action of the Hamiltonian on the vacuum. 


\subsection{Time-dependent functions $f_{nl}$ and $g_{nl}$}

Let us allow now that the functions $f_{nl}$ and $g_{nl}$ depend on time. Under the corresponding time-dependent canonical transformation, the Hamiltonian changes according to
\begin{equation}\label{4.11}
-\tilde{\sum_{nlm}} \tilde{\Pi}_{nl m} \tilde{\phi}_{nl m}' + H = -i\tilde{\sum_{nlm}} a^{*}_{nl  m}a'_{nl m} + \mathring{H}.
\end{equation}
In terms of the new canonical variables, the old Hamiltonian $H$ and the Legendre term $-\tilde{\Pi}_{nl  m} \tilde{\phi}'_{nl  m}$ read, up to total derivatives, as follows
\begin{equation} \label{4.12}
\begin{aligned}
   H = \tilde{\sum_{nlm}} \left( - \frac{b_{\l}}{2} \right) &\Bigl\{ a^{*}_{nl  m}a^{*}_{nl  m} \left[f_{nl }^2 + g_{nl }^2 \left(k^2 + s_{\l} \right) \right] + a_{nl  m}a_{nl  m} \left[f_{nl }^{*2} + g_{nl }^{*2} \left(k^2 + s_{\l} \right) \right] \\ &-2a^{*}_{nl  m}a_{nl m} \left[ |f_{nl }|^2 + |g_{nl }|^2 \left(k^2 + s_{\l} \right) \right]\Bigr\},
\end{aligned}
\end{equation}
\begin{equation}\label{4.13}
\begin{aligned}
    -\tilde{\Pi}_{nl  m} \tilde{\phi}'_{nl  m} &= -i a^{*}_{nl  m}a'_{nl  m} + 
    \frac{1}{2}(f^{*'}_{nl }g^{*}_{nl }-f^{*}_{nl }g^{*'}_{nl }) a_{nl m}a_{nl m} \\
    &+ \frac{1}{2} (f'_{nl }g_{nl }-f_{nl }g'_{nl })a_{nl  m}^{*}a_{nl  m}^{*} + 
    (f_{nl }g_{nl }^{*'}-f_{nl }^{*'}g_{nl }) a_{nl m}^{*}a_{nl m} .
    \end{aligned}
\end{equation}
Using this relation and the derivative of Eq. \eqref{2.11}, we can write
\begin{equation}\label{4.14}
\begin{aligned}
       \mathring{H} &= H + \frac{1}{2} \tilde{\sum_{nlm}} \Bigl\{ 2 \mathscr{R} \left\lbrace g'_{nl  }f^{*}_{nl  }-f'_{nl  }g^{*}_{nl } \right\rbrace a^{*}_{nl m} a_{nl m}\\ 
        &+ (f'_{nl } g_{nl }-f_{nl }g'_{nl  })a^{*}_{nl  m}a^{*}_{nl m} + (f^{*'}_{nl  }g^{*}_{nl  }-f^{*}_{nl }g^{*'}_{nl }) a_{nl  m}a_{nl m} \Bigr\}.
    \end{aligned}
\end{equation}

Now, let us consider the action on the vacuum state. Imposing normal ordering, the condition for a well-defined action becomes
\begin{equation}\label{4.15}
    \bra{0} \hat{\mathring{H}}^\dagger \hat{\mathring{H}} \ket{0} = \tilde{\sum_{nlm}} \frac{b^2_{\l}}{2} \ \Big| f_{nl }^2 + g_{nl }^2(k^2+s_{\l})-\frac{1}{b_{\l}}(f'_{nl }g_{nl }-f_{nl }g'_{nl }) \Big|^2 < \infty.
\end{equation}
Substituting $f_{nl}$ and $g_{nl}$ from Eq. \eqref{2.14} and neglecting subdominant terms with respect to the $k-$dependence, we get
\begin{equation}\label{4.16}
    \begin{aligned}
        \tilde{\sum_{nlm}} \ \frac{b^2_{\l}}{2} &\bigg| \left[\sqrt{2k^3} +\frac{ik}{b_{\l}}\vartheta^{(f)'}_{nl }\right] \left( \vartheta^{(f)}_{nl } +i \vartheta^{(g)}_{nl } \right) - \frac{s_{\l}}{2k} - \frac{i}{b_{\l}} \sqrt{\frac{k}{2}} \left( \vartheta^{(f)'}_{nl } +i \vartheta^{(g)'}_{nl } \right) \bigg|^2 < \infty.
    \end{aligned}
\end{equation}
We will restrict ourselves to functions $f_{nl }$ and $g_{nl }$ for which the time derivative respects the asymptotic order of their expansions for large $k$. As a consequence, we have $\vartheta^{(f)'}_{nl } +i \vartheta^{(g)'}_{nl } =o(k^{1/2})$. In addition, $\vartheta^{(f)'}_{nl } = o(k^{1/2})$, so that, when multiplied by $k$,  it is negligible compared to $\sqrt{2k^3}$. Therefore, recalling that the functions $\vartheta^{(f)}_{nl }$ and $ \vartheta^{(g)}_{nl }$ can now depend on time, similar arguments to those employed above let us conclude that we must have
\begin{equation}\label{4.18}
\vartheta^{(f)}_{nl} +i \vartheta^{(g)}_{nl} = \frac{s_{l/N}}{\sqrt{8k^5}} + \Gamma_{nl},
\end{equation}
where $\Gamma_{nl}$ is $o(k^{-5/2})$. Note that, at this order, $s_{\l}k^{-5/2}$ is equivalent to $s_{l/N}k^{-5/2}$.

In Eq. \eqref{4.16} there is then a term $\sqrt{2k^3}\Gamma_{nl}$ which behaves asymptotically as $o(k^{-1})$. There are other subdominant terms that we have ignored to arrive at Eq. \eqref{4.16} which may also be of order $o(k^{-1})$. The summability of all these remaining terms may still impose conditions on $\Gamma_{nl}$, restricting some of its leading-order asymptotic contributions. 

The crucial point is that we opened up the possibility to overcome the ill-definition of the Hamiltonian action. In fact, the obstruction put forward in the previous section concerned the strict dynamics of the canonical pair $(\tilde{\phi}_{nlm}, \tilde{\Pi}_{nlm})$, which can be obviated, as we have shown, by a modified dynamics introduced via the time dependence of the  representation functions $f_{nl}$ 
and $g_{nl}$.
 
However, this introduces back additional freedom in our description of the system, and therefore extra criteria are required in order to remove this ambiguity.  In the homogeneous case, an asymptotic diagonalization of the Hamiltonian in the ultraviolet regime \cite{BGT} has been proposed for this purpose. This proposal has indeed proven successful in removing the undesired extra freedom in the quantization, selecting in particular a unique vacuum. In the following, we will demonstrate that this method can also be applied to our anisotropic case. We will also show how to express the Klein-Gordon field in terms of functions that depend on the background and creation and annihilation variables with unitary Heisenberg dynamics.

\subsection{Hamiltonian diagonalization}

Although we have reached a Hamiltonian with a well-defined action on the vacuum and the corresponding particle states, we can note that this Hamiltonian in principle creates and destroys infinite pairs of particles. This arises from an infinite linear combination of self-interaction terms within the Hamiltonian which do not leave any domain of finite-particle states invariant. With this question in mind, we now want to address whether the still available freedom in the choice of $\Gamma_{nl}$ may allow a diagonalization (at least asymptotically) of the Hamiltonian, so that the corresponding operator becomes (asymptotically) proportional to the number operator. Indeed, we will show that there is no obstruction to applying the asymptotic diagonalization procedure presented in Ref. \cite{BGT} to our anisotropic model. 

Let us impose that the self-interaction terms in Eq. \eqref{4.14} be zero in the asymptotic regime of large $k$. In terms of the variables introduced in Eq. $\eqref{4.4}$, this condition reads
\begin{equation}\label{5.1}
    -ib_{\l}\f\xi_{nl} + \frac{\b\f^2\s}{2k^2}- \frac{\f\x'}{2k}+\frac{\f'\x}{2k}-\frac{i\b\f\x\s}{k^2}-\frac{\b\x^2}{2}-\frac{\b\x^2\s}{2k^2}=0.
\end{equation}
Proceeding as in Ref. \cite{BGT}, we can solve this equation order by order. We first impose that the function $-\xi_{nl}$ must be equal to the dominant-order terms (so that they cancel each other), plus a subdominant function. In this way, we find an algorithm to determine $\xi_{nl}$ as an asymptotic series, namely
\begin{equation} \label{5.2}
   \x= -\frac{i\f}{2k^2} \sum_{j}^{\infty} \left[ \frac{-i}{2k}\right]^j \gamma_j,
\end{equation}
where $\gamma_j$ are functions of the mass $\s$ and its derivatives. Computations similar to those explained in Ref. \cite{BGT} lead to\footnote{We notice that, if the background also adopts a Hamiltonian description, the time derivatives (denoted by a prime) coincide with Poisson brackets on the background phase space. This (when applied to Eq. \eqref{5.3} below) implies that the time dependence of the functions $\gamma_j$ can be expressed directly only in terms of background phase space variables.} 
\begin{equation}\gamma_0 = \s,\quad \gamma_1 = -\frac{\gamma_0^{\prime}}{\b}, \quad \gamma_2 = -\frac{\gamma_1^{\prime}}{\b}+3\gamma_0^2, \quad \gamma_3 = -\frac{\gamma_2^{\prime}}{\b} + 2\gamma_0\gamma_1,\quad {\rm etc.}
\end{equation} 
Actually, we can determine each $\gamma_j$ by the following formula
\begin{equation}\label{5.3}
    \gamma_{j+1} = -\frac{1}{\b} \gamma_{j}^{\prime} + 4\s \left[ \gamma_{j-1} + \sum_{r=0}^{j-3} \gamma_{r} \gamma_{j-(r+3)}\right]- \sum_{r=0}^{j-1} \gamma_{r} \gamma_{j-(r+1)}, \indent \forall j \geq 0,
\end{equation}
and zero for every $\gamma_j$ with $j<0$. Using Eqs. \eqref{2.14} and \eqref{4.4}, we can check that $\vaf + i \vag = \frac{i}{k}\x$. Hence, our recurrence equation and the definition of $\x$ allow us to fix $\vag$ in terms of $\vaf$. Explicitly,
\begin{equation}\label{5.4}
    \vag = -\frac{i}{\sqrt{8k^5}} \sum_{j}^{\infty} \left( \frac{-i}{2k}\right)^j \gamma_j + \frac{i}{2k^2} \vaf
 \left[2k^2-\sum_{j}^{\infty} \left( \frac{-i}{2k}\right)^j \gamma_j \right].
 \end{equation}
However, $\vaf$ still remains to be determined. We will achieve this by first fixing its modulus and then its phase, or what is equivalent, by fixing the modulus and phase of $\f$, since its relation to $\vaf$ is direct. 

Starting from Eq. \eqref{2.11} and employing Eqs. \eqref{4.4} and \eqref{5.2}, we arrive at
\begin{equation}\label{5.5}
    |\f|^2 \left[ 1- \frac{1}{2k^2} \sum_{j=0}^{\infty} \left( \frac{i}{2k}\right)^{2j} \gamma_{2j}\right] = \frac{k}{2}.
\end{equation}
This equation determines the modulus of $\f$ once the functions $\gamma_{j}$ have been found by our recurrence procedure.

On the other hand, to fix the phase of $\f$, it is convenient to introduce the function $\h=g^{-1}_{nl} f_{nl}$. Notice that, thanks to Eq. \eqref{4.3}, $\g \neq 0$ and hence $\h$ is well defined \cite{BGT}. In terms of $\h$, condition \eqref{5.1} reads $\h' = \b(k^2+s_{\l}+h^2_{nl})$. The freedom in choosing $\f$ and $\g$ can be seen as an ambiguity in determining a unique solution to this last differential equation. In fact, given a solution $\h$ we can calculate (if existing) $k h^{-1}_{nl}-i= \f^{-1}\x$ as an asymptotic expansion in terms of $\gamma_j$. Note from Eq. \eqref{2.11} that $2\mathscr{I}\{\h\} = -|\g|^{-2}$, i.e. $\mathscr{I}\{\h\}$ should be a negative function. Besides, then we can write $2|\f|^2=-|\h|^2\mathscr{I}^{-1}\{\h\}$. 

With these preliminaries, we can now proceed to fix the phase of $\f$, which we call $F_{nl}$. It is not difficult to see that, leaving apart its zero mode, the scalar field $\Phi$ can be expressed in terms of our creation and annihilation variables as
\begin{equation} \label{5.9}
    \begin{aligned}
        \Phi &=i\tilde{\sum_{nlm}}\frac{1}{\sqrt{\b(\eta_{\l})}}[\g^{*}a_{nlm}(\eta_{\l})-\g a^{*}_{nlm}(\eta_{\l})]R_n(r) \Bar{Y}_l^m(\theta,\varphi) \\
    &= \tilde{\sum_{nlm}} \frac{|\f(\eta_{\l})|}{k\sqrt{\b(\eta_{\l})}} \left[ 1- \frac{1}{2k^2} \sum_{j=0}^{\infty} \left( \frac{i}{2k}\right)^j \gamma_j(\eta_{\l}) \right] a_{nlm}(\eta_{\l}) e^{-iF_{nl}(\eta_{\l})} R_n(r) \Bar{Y}_l^m(\theta,\varphi) + \text{h.c.},
    \end{aligned}
\end{equation}
where the last term denotes Hermitian conjugation, and $R_n$ and $\Bar{Y}_l^m$ are, respectively, real Fourier modes and real spherical harmonics. Explicitly, the Fourier modes are
\begin{eqnarray}
R_n(r) &=& \frac{1}{\sqrt{\pi }} \sin\left(nr\right)  \quad {\rm for} \quad n<0, \nonumber\\ 
R_n(r) &=& \frac{1}{\sqrt{\pi}} \cos\left(nr\right) \quad {\rm for} \quad n>0,
\end{eqnarray}
with $R_0=1/\sqrt{2\pi}$, while the real spherical harmonics are
\begin{eqnarray}
\bar{Y}_l^m (\theta, \varphi) &= &\frac{Y_l^m(\theta, \varphi)  +(-1)^mY_l^{-m}(\theta, \varphi) }{\sqrt{2}}, \quad {\rm for} \quad m>0, \nonumber\\
\bar{Y}_l^m (\theta, \varphi) &=& i \frac{Y_l^m(\theta, \varphi)  -(-1)^mY_l^{-m}(\theta, \varphi) }{\sqrt{2}}, \quad {\rm for} \quad m<0,
\end{eqnarray} 
with $\bar{Y}_l^0 =Y_l^0$ and $Y_l^m$ denoting the complex spherical harmonics
\begin{equation}
Y^m_l(\theta, \varphi) = (-1)^{m} \sqrt{\frac{(2l+1)}{4\pi} \frac{(l-m)!}{(l+m)!}} P^{m}_l(\cos(\theta)) e^{im\varphi}.
\end{equation}
Here, $P^{m}_l$ are the Legendre polynomials. Note that the complex conjugate of $Y_l^m$ is $(-1)^m Y_l^{-m}$. 

On the other hand, one can check that the (asymptotically) diagonalized Hamiltonian can be written in the form \cite{BGT}
\begin{equation}\label{5.8}
    \mathring{H} = \tilde{\sum_{nlm}} \b  \Lambda_{nl}  a^{*}_{nlm} a_{nlm}, \indent \text{ with } \ \Lambda_{nl} = -\frac{F_{nl}^{\prime}}{\b}  - \frac{k^2+\s}{|\h|^2} \mathscr{I} \{ \h \}.
\end{equation}
Solving then the Hamiltonian equations of motion in conformal time $\eta_{\l}$ and noting that $\Lambda_{nl}$ must be a purely real function, we get
\begin{equation} \label{5.10}
    \begin{pmatrix}
        a_{nlm}(\eta_{\l})\\
        a^{*}_{nlm}(\eta_{\l})
    \end{pmatrix} = \begin{pmatrix}
        {E (\eta_{\l},\eta^{\circ}_{\l})}  & 0 \\
        0 &  {E^{*}(\eta_{\l},\eta^{\circ}_{\l})}
    \end{pmatrix} \begin{pmatrix}
        a_{nlm}(\eta^{\circ}_{\l})\\
        a^{*}_{nlm}(\eta^{\circ}_{\l})
    \end{pmatrix}, \quad {E(\eta_{\l},\eta^{\circ}_{\l})=\exp\left[-i \int_{\eta^{\circ}_{\l}}^{\eta_{\l}} d\eta_{\l}' \Lambda_{nl}(\eta'_{\l})\right],}
\end{equation}
where $\eta^{\circ}_{\l}$ is an arbitrary initial time. Thus, the (asymptotic) Heisenberg evolution of the creation and annihilation variables amounts to a time-dependent phase. We could choose $F_{nl}$ to absorb all the dynamical phase of $a_{nlm}$ and $a^{*}_{nlm}$ in Eq. \eqref{5.9}, but this would remove all the physical information concerning to the dynamics. On the contrary, the time dependence of $\f$ (and $\g$) can be chosen to balance only in norm that part of the field $\Phi$ that displays a direct dependence on the evolving background (the part in square brackets and the $\b$ global factor). In this way, as proposed in Ref. \cite{BGT}, we respect as much as possible the dynamical behavior of the system while permitting a unitary evolution. In this spirit, and once realized that $k^{-1}|\f|/\sqrt{\b}$ is a purely real function, we fix the phase of $\f$ as follows:
\begin{equation}\label{5.11}
    F_{nl}(\eta_{\l}) = \arg \left[ 1 - \frac{1}{2k^2} \sum_{j=0}^{\infty} \left(\frac{i}{2k}\right)^j \gamma_{j}(\eta_{\l})\right].
\end{equation}

With this choice of $F_{nl}$, the (asymptotic) frequency appearing in the diagonalization of the Hamiltonian is \cite{BGT} $\Lambda_{nl} = - \mathscr{I} \{ \h \}$, which should be positive according to our comments above. With the help of Eqs. \eqref{4.4} and \eqref{5.2}, we can rewrite it as
\begin{equation}\label{5.15}
    \Lambda_{nl} = \mathscr{R} \left\lbrace \frac{k}{1-\frac{1}{2k^2}\sum_{j} \left[\frac{-i}{2k}\right]^j \gamma_{j}}\right\rbrace.
\end{equation}
Correspondingly, the mode expansion of the scalar field (ignoring the zero mode and omitting the time dependence of $\gamma_j$ and $b_{\hat{l}}$ to simplify the notation) becomes 
\begin{equation}\label{5.14}
    \Phi= \tilde{\sum_{nlm}} \frac{\Big| 1- \frac{1}{2k^2}\sum_{j} \left( \frac{i}{2k}\right)^j \gamma_{j} \Big|}{\sqrt{1-\frac{1}{2k^2} \sum_{j} \left( \frac{i}{2k}\right)^{2j} \gamma_{2j}}} \frac{e^{-i \int d\bar{\tau} \b \Lambda_{nl}}}{\sqrt{2k\b}} a_{nlm}(\tau_0) R_n(r) \bar{Y}_{l}^{m}(\theta,\varphi) + \text{h.c.}
\end{equation}
With a little bit of algebra, and using Eq. \eqref{5.15}, the integral in the above exponential factor can be expressed as
\begin{equation}
    \begin{aligned}
       & - i \int_{\tau_0}^{\tau} \frac{d\bar{\tau} \b \left(k-\frac{1}{2k}\sum_j (-1)^j\left[ \frac{1}{2k}\right]^{2j} \gamma_{2j}\right)}{1-\frac{1}{k^2}\sum_j (-1)^j\left[ \frac{1}{2k}\right]^{2j} \gamma_{2j}+ \frac{1}{4k^4}\left[ \sum_{j,r} (-1)^{j+r} \left[ \frac{1}{2k}\right]^{2(j+r)} \gamma_{2j}\gamma_{2r} + \sum_{j,r}(-1)^{j+r} \left[\frac{1}{2k}\right]^{2(j+r+1)} \gamma_{2j+1}\gamma_{2r+1}\right]}. \nonumber
    \end{aligned}
\end{equation}

Let us remark that, since the Hamiltonian \eqref{5.8} is diagonal, our beta-function in Eq. \eqref{2.12} is zero (in the asymptotic regime), so that it is square summable. Thus, there is no need for further constraints to guarantee the square summability. The Hamiltonian diagonalization imposes enough restrictions on the system: it straightforwardly allows for a unitary implementation in the asymptotic sector. Finally, note that, except for a phase that depends on the integral of the frequency $\Lambda_{nl}$ over time (which nonetheless could be transformed into an integral over a phase space variable with a suitable change from the coordinate time $\tau$ to an internal time), the time dependence in our final expression for the scalar field is purely based on the dependence on the background metric functions via $\b$ and the mass $\s$ and its derivatives (see Eq. \eqref{5.3}).

\section{Conclusions} \label{Co}

Invariance of the vacuum under spatial isometries and the requirement of a unitary Heisenberg dynamics are capable of selecting a family of unitarily equivalent Fock representations for a scalar field. This result is valid not only in isotropic cosmologies, but also in anisotropic ones like the case considered here of a Kantowski-Sachs spacetime \cite{Prev}. This type of spacetimes has received a lot of attention recently, since they can describe the interior geometry of nonrotating black holes \cite{AOS,Andres}. In this work, we have shown that the criteria of spatial symmetries and unitary dynamics determine a unique field redefinition by means of anisotropic rescalings, which respect the invariance under spatial isometries and depend only on time and on the anisotropic label $\hat{l}$ of the field modes in a decomposition in terms of LB eigenstates. This result extends similar results for isotropic scenarios \cite{CMV,PRD83}. 

One could then adopt the massless representation, as one of the Fock representations in the family selected by these rescalings, and restrict all considerations to equivalent representations obtained by means of time-independent canonical transformations. For certain purposes, a quantum field theory constructed in this way can be sufficient since, in particular, it would display a unitary Heisenberg evolution. Nonetheless, if we want a quantum Hamiltonian with nice physical properties, we have to go one step beyond and allow for time-dependent canonical transformations. Actually, we have shown that this generalization is necessary if we want the Hamiltonian to have a well-defined action on the vacuum. Time-dependent canonical transformations assign part of the field evolution to the background, so that the remaining dynamical evolution of the creation and annihilation variables can have a much better behavior.

As a way to completely specify the functions $\f$ and $\g$ that parametrize the canonical transformation, and hence fix all the existing ambiguity in our choice of Fock representation and vacuum state, we have proposed to introduce an asymptotic Hamiltonian diagonalization along the lines of Ref. \cite{BGT}. We have proven that there are no obstructions to this program, and that it entirely determines the creation and annihilation variables (at least asymptotically). In this asymptotic diagonalization, the phase of the functions $\f$ are chosen so as to avoid that they absorb dynamical information. With this extra requirement, no freedom is left in the construction. One obtains a privileged splitting of the scalar field evolution between a dependence on the background and a Heisenberg dynamics for the creation and annihilation variables. It is these last dynamics which admit a unitary implementation and have an associated Hamiltonian that is asymptotically diagonal. These dynamics are solvable, and therefore it is possible to achieve an explicit expression of the Klein-Gordon field in which the intricacy of the time variation is captured on the background dependence. 

The results obtained in the present article pave the way to explore the application of the hybrid approach to LQC for black hole systems with field content or with the inclusion of perturbations \cite{Andres}. It would also be of the greatest interest to study the extension of the geometry and the scalar quantum field to the exterior of the black hole. One would then be in a good position to compare and discuss the selected non-oscillating vacuum with other proposed vacua for scalar fields in black holes spacetimes \cite{Wald}. In addition, it could open the door to a notion of unitary transformations for black holes with geometric significance.

\acknowledgments
This work was partially supported by Project No. MICINN PID2020-118159GB-C41 from Spain. J.V.\ is grateful for the support given by the research unit Fiber Materials and Environmental Technologies (FibEnTech-UBI), on the extent of the project reference UIDB/00195/2020, funded by the Funda\c c\~ao para a Ci\^encia e a Tecnologia (FCT), IP/MCTES through national funds (PIDDAC).  The authors are grateful to Beatriz Elizaga Navascu\'es and Andr\'es M\'{\i}nguez-S\'anchez for discussions.

\end{document}